# A distributed regression analysis application based on SAS software

# Part II: Cox proportional hazards regression

**Authors**

**Yury Vilk, PhD**
Department of Population Medicine
Harvard Medical School and Harvard Pilgrim Health Care Institute
401 Park Street, Suite 401 East
Boston, MA 02215
Yury_Vilk@harvardpilgrim.org
617-867-4891

**Zilu Zhang, MSc**
Department of Population Medicine
Harvard Medical School and Harvard Pilgrim Health Care Institute
401 Park Street, Suite 401 East
Boston, MA 02215
Zilu_Zhang@harvardpilgrim.org

**Jessica Young, PhD**
Department of Population Medicine
Harvard Medical School and Harvard Pilgrim Health Care Institute
401 Park Street, Suite 401 East
Boston, MA 02215
jyoung@hsph.harvard.edu

**Qoua L. Her, PharmD, MSPharm, MSc**
Department of Population Medicine
Harvard Medical School and Harvard Pilgrim Health Care Institute
401 Park Street, Suite 401 East
Boston, MA 02215
Qoua_Her@harvardpilgrim.org
617-867-4885

**Jessica M. Malenfant, MPH**
Department of Population Medicine
Harvard Medical School and Harvard Pilgrim Health Care Institute
401 Park Street, Suite 401 East
Boston, MA 02215
Jessica_Malenfant@harvardpilgrim.org

**Sarah Malek, MPPA**

Department of Population Medicine
Harvard Medical School and Harvard Pilgrim Health Care Institute
401 Park Street, Suite 401 East
Boston, MA 02215
Sarah_Malek@harvardpilgrim.org

**Sengwee Toh, ScD**
Department of Population Medicine
Harvard Medical School and Harvard Pilgrim Health Care Institute
401 Park Street, Suite 401 East
Boston, MA 02215
Darren_Toh@harvardpilgrim.org

**Word Count**:

# ABSTRACT

Previous work has demonstrated the feasibility and value of conducting distributed regression analysis (DRA), a privacy-protecting analytic method that performs multivariable-adjusted regression analysis with only summary-level information from participating sites. To our knowledge, there are no DRA applications in SAS, the statistical software used by several large national distributed data networks (DDNs), including the Sentinel System and PCORnet. SAS/IML is available to perform the required matrix computations for DRA in the SAS system. However, not all data partners in these large DDNs have access to SAS/IML, which is licensed separately. In this second article of a two-paper series, we describe a DRA application developed using Base SAS and SAS/STAT modules for distributed Cox proportional hazards regression within horizontally partitioned DDNs and its successful tests.

*Word Count (max. 250): 124*

**Funding Statement:** This work was supported by the Office of the Assistant Secretary for Planning and Evaluation (ASPE) and the Food and Drug Administration (HHSF223201400030I/HHSF22301006T).

**Competing Interests Statement:** Dr. Toh is Principal Investigator of projects funded by the National Institutes of Health (U01EB023683) and the Patient-Centered Outcomes Research Institute (ME-1403-11305).

**Contributorship Statement:** All authors contributed to the conception, design, analysis, and interpretation of this study. YV, JY, and ST led the drafting of the manuscript and revising it for critical important intellectual content. QLH, JMM, SM, and ZZ contributed to the conception and design, drafting of the manuscript, or revising it. All authors commented on manuscript drafts and gave their approval for the final version to be published. ST obtained funding and was responsible for supervision of all activities.

# 1. Introduction

Sharing of detailed individual-level information raises concerns about individual privacy and confidentiality, which may deter multi-center collaborations (Maro et al. 2009, Brown et al. 2010, Toh et al. 2011). Data organized in a distributed data network (DDN), where data remain behind each data partner's firewall, alleviates some of these concerns (Diamond, Mostashari, and Shirky 2009, Maro et al. 2009, Brown et al. 2010, Toh et al. 2011). Distributed regression analysis (DRA) within a DDN is one approach that can help overcome privacy concerns, allowing multivariable regression analysis using only summary-level information and producing equivalent results to those from pooled individual-level data analysis (Karr et al. 2004, Fienberg et al. 2006, Wolfson et al. 2010, Wu et al. 2012, Toh et al. 2014, Dankar 2015).

There are currently several R-based software applications that allow users to perform DRA for the Cox proportional hazards regression model (Wolfson et al. 2010, Jiang et al. 2013, Lu et al. 2015, Meeker et al. 2015, Narasimhan et al. 2017). To our knowledge, there are no Cox DRA applications in SAS, the statistical software used by several existing national DDNs in the United States, including the Sentinel System (a DDN funded by the U.S. Food and Drug Administration to conduct medical product safety surveillance) (Platt et al. 2012, Ball et al. 2016) and PCORnet (a DDN funded by the Patient-Centered Outcomes Research Institute to perform comparative effectiveness research) (Fleurence et al. 2014).

In our companion paper, we describe a DRA application for linear and logistic DRA (Her et al. Submitted). Here, we describe a DRA application for Cox proportional hazards regression within a horizontally partitioned DDN, a data environment in which different databases include information from different individuals. The application comprises two interlinked packages of macros and programs – one for the analysis center and one for the data-contributing sites (i.e., data partners). As in the case of our linear and logistic DRA application (Her et al. Submitted), a key advantage of our Cox DRA application is that it requires only Base SAS and SAS/STAT modules, avoiding reliance on SAS/IML, which is licensed separately. Another advantage is that we have fully integrated the DRA application with PopMedNet$^{TM}$, an open-source query distribution software application that supports automatable file transfer between the analysis center and data partners (Her et al. 2018).

We organize the article as follows. In Section 2, we describe our implementation of the DRA algorithm for the Cox model using only Base SAS and SAS/STAT to compute parameter estimates, standard errors, and goodness-of-fit measures. Our application handles tied events using either the Breslow or Efron approximation and can fit either a non-stratified or stratified Cox model. In Section 3, we briefly summarize steps involved in setting up and executing the DRA application for Cox proportional hazards regression. Many of the steps are shared across linear, logistic, and Cox DRA, so we only summarize the common steps in this article, highlight what is different for Cox DRA, and refer readers to the companion paper that describes these steps in greater detail (Her et al. Submitted). In Section 4, we present results from the Cox DRA application in two empirical examples and compare them with the results obtained from standard SAS procedures that analyzed pooled individual-level datasets. In Section 5, we discuss possible extensions of the Cox DRA application.

## 2. Distributed Cox regression analysis for horizontally partitioned data

### 2.1 Overview

We consider a horizontally partitioned DDN where each data partner holds distinct individual cohorts. Our implementation uses a secure protocol with a semi-trusted third party as the analysis center as described in our companion paper (Her et al. Submitted). Below we describe our computational algorithm to implement Cox DRA using only the Base SAS and SAS/STAT modules.

### 2.2 Computational algorithm

In this section, we describe the underlying algorithm implemented by our DRA application, which is a distributed version of the Newton-Raphson algorithm implemented to solve for the parameter estimates of a stratified Cox model based on the Breslow approximation to the partial likelihood for tied event times (Breslow 1974). The algorithm fits a non-stratified Cox model when the number of strata is set to 1. This algorithm avoids sharing individual-level data by a given data partner with other sites and with the analysis center. As described in **Appendix A**, our algorithm can also implement the Efron approximation (Efron 1977).

For $k = 1, \ldots, K$, $m = 1, \ldots, M$, and $i = 1, \ldots, N_{m,k}$, let $K$ denote the number of sites, $M$ the number of strata, and $N_{m,k}$ the number of subjects at site $k$ in strata $m$. Suppose, among all $N_m = \sum_{k=1}^{K} N_{m,k}$ patients in strata $m$, there are $J_m$ unique event times, $t_{m,1} < t_{m,2} < \cdots < t_{m,J_m}$. Denote $(w_{i,m,k}, T_{i,m,k}, \Delta_{i,m,k}, \mathbf{Z}_{i,m,k})$ as the observed data for subject $i$ at site $k$ in stratum $m$, with $T_{i,k,m}$ representing the observed follow-up time, $\Delta_{i,m,k}$ the censoring indicator (1 if $T_{i,m,k}$ corresponds to the event time and 0 if the censoring time), $w_{i,m,k}$ an individual-level weight and $\mathbf{Z}_{i,m,k}$ a $p * 1$ vector of covariates. Define $d_{m,j} = \sum_{k=1}^{K} \sum_{i=1}^{N_{m,k}} I(T_{i,m,k} = t_{m,j}, \Delta_{i,m,k} = 1)$ as the number of events at time $t_{m,j}$ from all sites. Here the function $I(a = b, c = d, \ldots)$ is defined to be equal to 1 when all conditions are true and 0 otherwise.

The input dataset at site $k$ has the following structure for stratum $m$:

$$
\begin{array}{cccccc}
w_{1,m,k} & T_{1,m,k} & Z_{1,m,k,1} & \cdots & Z_{1,m,k,p} & \Delta_{1,m,k} \\
\vdots & \vdots & \vdots & & \vdots & \vdots \\
w_{N_{m,k},m,k} & T_{N_{m,k},m,k} & Z_{N_{m,k},m,k,1} & \cdots & Z_{N_{m,k},m,k,p} & \Delta_{N_{m,k},m,k}
\end{array} \quad (1)
$$

Under a stratified Cox model, the hazard function for subjects at site $k$ within stratum $m$ for covariate level $\mathbf{Z}_{i,m,k}$ is assumed to have the following form:

$$h_m(t|\mathbf{Z}_{i,m,k}) = \exp(\boldsymbol{\beta}^T \mathbf{Z}_{i,m,k}) \, h_m^{(0)}(t) \quad (2)$$

where $\boldsymbol{\beta}$ is an unknown $p * 1$ vector of regression coefficients. In the special case of $M = 1$, the model in Equation (2) reduces to a non-stratified Cox model. Another important special case occurs when data partner identifier is one of the stratification variables. We consider this case further below.

We use a Newton-Raphson algorithm to calculate the partial likelihood estimator of the regression coefficients $\boldsymbol{\beta}$. To apply this algorithm in DDNs, the log-likelihood $l(\boldsymbol{\beta})$, gradient $\boldsymbol{g}(\boldsymbol{\beta}) = \frac{\partial l(\boldsymbol{\beta})}{\partial \boldsymbol{\beta}}$, and the Hessian matrix $\boldsymbol{H}(\boldsymbol{\beta}) = \frac{\partial^2 l(\boldsymbol{\beta})}{\partial \boldsymbol{\beta} \partial \boldsymbol{\beta}^T}$ must be expressed in terms of aggregated quantities from each data partner. Let's first define the quantities that have to be calculated at each site $k$ in each stratum $m$.

Define:

$$d_{m,j,k}^{(l)} = \sum_{i=1}^{N_{m,k}} I(T_{i,m,k} = t_{m,j}, \Delta_{i,m,k} = 1) w_{i,m,k} Z_{i,m,k}^l \quad (3)$$

$$S_{m,j,k}^{(l)}(\boldsymbol{\beta}) = \sum_{i=1}^{N_{m,k}} I(T_{i,m,k} \geq t_{m,j}) w_{i,m,k} \exp(\boldsymbol{\beta}^T Z_{i,m,k}) Z_{i,m,k}^l \quad (4)$$

Where $l = 0,1$ for $d_{m,j,k}^{(l)}$ and $l = 0,1,2$ for $S_{m,j,k}^{(l)}(\boldsymbol{\beta})$, $Z_{i,m,k}^0 = 1$, $Z_{i,m,k}^1 = Z_{i,m,k}$, $Z_{i,m,k}^2 = Z_{i,m,k} Z_{i,m,k}^T$ such that $d_{m,j,k}^{(l)}$ and $S_{m,j,k}^{(l)}(\boldsymbol{\beta})$ are scalars when $l = 0$, vectors with dimension $p$ when $l = 1$ and $S_{m,j,k}^{(2)}(\boldsymbol{\beta})$ is a matrix with dimensions $p * p$ when $l = 2$.

Below we use a notation in which an absence of an index in a matrix implies summation over that index. For example,

$$d_{m,j}^{(l)} = \sum_k d_{m,j,k}^{(l)} \qquad\qquad S_{m,j}^{(l)} = \sum_k S_{m,j,k}^{(l)} \quad (5)$$

Note, that when the list of stratification variables (index $m$) includes a data partner identifier represented by index $k$, the summation over $k$ does not change the results in Equation (5) because there is only one possible value of $k$ at a given $m$. However, in the more general case, summation over site index $k$ is necessary.

The log-likelihood $l(\boldsymbol{\beta})$, gradient $\boldsymbol{g}(\boldsymbol{\beta})$ and the Hessian matrix $\boldsymbol{H}(\boldsymbol{\beta})$ can be written in terms of these summarized quantities:

$$l(\boldsymbol{\beta}) = \sum_{m=1}^{M} l_m(\boldsymbol{\beta}) \qquad \boldsymbol{g}(\boldsymbol{\beta}) = \sum_{m=1}^{M} \boldsymbol{g}_m(\boldsymbol{\beta}) \qquad \boldsymbol{H}(\boldsymbol{\beta}) = \sum_{m=1}^{M} \boldsymbol{H}_m(\boldsymbol{\beta}) \quad (6)$$

where

$$l_m(\boldsymbol{\beta}) = \sum_j \left\{ \boldsymbol{\beta}^T \boldsymbol{d}_{m,j}^{(1)} - d_{m,j}^{(0)} \log S_{m,j}^{(0)}(\boldsymbol{\beta}) \right\} \tag{7}$$

$$\boldsymbol{g}_m(\boldsymbol{\beta}) = \sum_j \left\{ \boldsymbol{d}_{m,j}^{(1)} - d_{m,j}^{(0)} \frac{\boldsymbol{S}_{m,j}^{(1)}(\boldsymbol{\beta})}{S_{m,j}^{(0)}(\boldsymbol{\beta})} \right\} \tag{8}$$

$$\boldsymbol{H}_m(\boldsymbol{\beta}) = -\sum_j d_{m,j}^{(0)} \left\{ \frac{\boldsymbol{S}_{m,j}^{(2)}(\boldsymbol{\beta})}{S_{m,j}^{(0)}(\boldsymbol{\beta})} - \frac{\boldsymbol{S}_{m,j}^{(1)}(\boldsymbol{\beta}) * \left[\boldsymbol{S}_{m,j}^{(1)}(\boldsymbol{\beta})\right]^T}{\left[S_{m,j}^{(0)}(\boldsymbol{\beta})\right]^2} \right\} \tag{9}$$

In general, these representations imply that the Newton-Raphson algorithm can be executed such that the summarized matrices $\boldsymbol{d}_{m,j,k}^{(l)}$ and $\boldsymbol{S}_{m,j,k}^{(l)}(\boldsymbol{\beta})$ are computed at each data partner site and transferred to the analysis center. The size of a dataset to store these matrices for all $j$ depends on the number of event times $J_m$ for stratum $m$. For example, the dataset to store all matrices $\boldsymbol{S}_{m,j,k}^{(2)}(\boldsymbol{\beta})$ for stratum $m$ has $p * p * J_m$ data elements. This can result in the need to transfer a significant amount of data when the number of event times is large. There may also be a significant risk from a privacy perspective because the number of observations contributing to the computations for each event time can be small.

However, the calculations can be done much more efficiently and with much smaller privacy risk when one stratifies on a set of variables that includes the data partner identifier. In multi-database studies, a stratified Cox model, stratified by data partner identifier, is more realistic than assuming a common baseline hazard function for all data partners. In this case, the summation over time event index $j$ in Equations (7) – (9) can be done at the data partners, resulting in the transfer of much smaller datasets to the analysis center. Specifically, only a dataset with matrices $l_m(\boldsymbol{\beta})$, $\boldsymbol{g}_m(\boldsymbol{\beta})$, and $\boldsymbol{H}_m(\boldsymbol{\beta})$, which are not dependent on the number of events times, need to be transferred to the analysis center. For example, the contribution to the Hessian matrix $\boldsymbol{H}_m(\boldsymbol{\beta})$ has dimension $p * p$.

The partial likelihood estimator of $\boldsymbol{\beta}$ is obtained by the Newton-Raphson algorithm, which on each iteration $n$ solves:

$$-\boldsymbol{H}(\boldsymbol{\beta}_n)(\boldsymbol{\beta}_{n+1} - \boldsymbol{\beta}_n) = \boldsymbol{g}(\boldsymbol{\beta}_n) \tag{10}$$

for $\boldsymbol{\beta}_{n+1}$ such that

$$\boldsymbol{\beta}_{n+1} = \boldsymbol{\beta}_n - \boldsymbol{H}^{-1}(\boldsymbol{\beta}_n)\boldsymbol{g}(\boldsymbol{\beta}_n) \tag{11}$$

based on an initial starting value $\boldsymbol{\beta}_0$ and iterating until a convergence criterion is met. Our goal is to solve these equations using only Base SAS and SAS/STAT modules as the SAS matrix manipulation module SAS/IML is licensed separately. From a computational perspective, the main challenge for solving Equation (10) for $\boldsymbol{\beta}_{n+1}$ is matrix inversion. Below we will illustrate how PROC REG (part of the SAS/STAT modules) can be used to solve the system of the Newton-Raphson linear Equation (10). Let us consider a system of linear equations

$$A\boldsymbol{b} = \mathbf{c} \tag{12}$$

where **A** is a symmetric, positive definite matrix with dimensions $p * p$, $c$ is a vector with dimension $p$ and $b$, an unknown coefficient vector. PROC REG can be used to solve a system of linear equations of the form of Equation (12) for $\boldsymbol{b}$ by passing in a sums of squares and cross products (SSCP) TYPE dataset in the form of:

$$\boldsymbol{SSCP} = \begin{pmatrix} \mathbf{A} & \mathbf{c}^T \\ \mathbf{c} & const \end{pmatrix} \tag{13}$$

When we pass a dataset in the form of Equation (13) into PROC REG, the solution to the system of linear equations of the form of Equation (12), is $\boldsymbol{b} = \mathbf{A}^{-1}\mathbf{c}$ which can be obtained by specifying the output dataset option in the PROC REG procedure. The diagonal element $const$ in row $p + 1$ and column $p + 1$ only affects the "regression" R-squared and has no effect on deriving $\boldsymbol{b}$. We use a very large number for this diagonal element ($const = 10^{12}$) to ensure that PROC REG does not produce a note in the log that R-squared is negative.

In our companion paper (Her et al. Submitted), we showed how this general capability of PROC REG with input dataset TYPE=SSCP can be used to implement linear regression and iteratively reweighted least squares for generalized linear models without the need for PROC IML. Although we cannot use iteratively reweighted least squares for Cox regression, we can exploit the capability of PROC REG to solve a system of linear equations. Specifically, the gradient vector $\boldsymbol{g}(\boldsymbol{\beta}_n)$ has length $p$ and the negative of the Hessian matrix $\mathbf{H} = -\mathbf{I}$ is symmetric and positive definite with dimension $* p$ , which is close to the partial likelihood estimate $\widehat{\boldsymbol{\beta}}$. Thus, we can use the above described approach to solve the Newton-Raphson Equation (10) by setting $\mathbf{A} = \mathbf{I}(\boldsymbol{\beta}_n) = -\mathbf{H}(\boldsymbol{\beta}_n)$ and $\mathbf{c} = \boldsymbol{g}(\boldsymbol{\beta}_n)$. For a given iteration $n$, the solution produced by PROC REG is:

$$\boldsymbol{b}_n = \mathbf{I}^{-1}(\boldsymbol{\beta}_n)\boldsymbol{g}(\boldsymbol{\beta}_n) \tag{14}$$

where $\boldsymbol{b}_n = \boldsymbol{\beta}_{n+1} - \boldsymbol{\beta}_n$. Using this solution, the next iteration of $\boldsymbol{\beta}^{n+1}$ can be computed as:

$$\boldsymbol{\beta}_{n+1} = \boldsymbol{\beta}_n + \boldsymbol{b}_n \tag{15}$$

In addition to coefficients $b_n$, PROC REG also outputs the inverse, $\mathbf{I}^{-1}$. The value of the matrix $\mathbf{I}^{-1}(\boldsymbol{\beta}) = -\frac{\partial^2 l(\boldsymbol{\beta})}{\partial \boldsymbol{\beta} \partial \boldsymbol{\beta}^T}$ evaluated at the final partial likelihood estimate $\boldsymbol{\beta} = \widehat{\boldsymbol{\beta}}$ gives us the estimated covariance matrix:

$$\widehat{cov}(\widehat{\boldsymbol{\beta}}) = \mathbf{I}^{-1}(\widehat{\boldsymbol{\beta}}) \tag{16}$$

Below we summarize our computational algorithm. The algorithm uses two different computational paths, which we refer to as *Case (a)* and *Case (b)*. *Case (a)* is implemented when the user specifies a stratified Cox model ($M > 1$) and the data partner identifier variable $dp\_cd$ is included in the list of stratification variables. *Case (b)* is implemented when the user specifies a non-stratified Cox model ($M = 1$), or a stratified Cox model ($M > 1$) but the data partner identifier variable $dp\_cd$ is not included in the list of stratification variables.

1) For each iteration $n + 1$ at each data partner site $k$, calculate matrices $\boldsymbol{d}_{m,j,k}^{(l)}(\boldsymbol{\beta}_n)$ and $\boldsymbol{S}_{m,j,k}^{(l)}(\boldsymbol{\beta}_n)$ using Equations (3) and (4) based on initial starting value $\boldsymbol{\beta}_0$.

   For *Case (a):* Calculate stratum-specific contributions to the log-likelihood $l_m(\boldsymbol{\beta}_n)$, gradient $\boldsymbol{g}_m(\boldsymbol{\beta}_n)$, and Hessian matrix $\boldsymbol{H}_m(\boldsymbol{\beta}_n)$ using Equations (7)- (9). In this case, these contributions can be calculated separately at each site and transferred to the analysis center, because the variable $dp\_cd$ is a stratification variable.

   For *Case (b)*: Bring matrices $\boldsymbol{d}_{m,j,k}^{(l)}(\boldsymbol{\beta}_n)$ and $\boldsymbol{S}_{m,j,k}^{(l)}(\boldsymbol{\beta}_n)$ from each site $k$ to the analysis center.

2) For each iteration $n + 1$ at the analysis center .

   For *Case (a)*: Sum the stratum-specific contributions $l_m(\boldsymbol{\beta}_n)$, $\boldsymbol{g}_m(\boldsymbol{\beta}_n)$, and $\boldsymbol{H}_m(\boldsymbol{\beta}_n)$ to obtain the log-likelihood $l(\boldsymbol{\beta}_n)$, gradient $\boldsymbol{g}(\boldsymbol{\beta}_n)$, and Hessian matrix $\boldsymbol{H}(\boldsymbol{\beta}_n)$ using Equation (6).

   For *Case (b)*: Sum contributions from all sites to obtain $\boldsymbol{d}_{m,j}^{(l)}$ and $\boldsymbol{S}_{m,j}^{(l)}$ using Equation (5). Calculate the stratum-specific contributions to the log-likelihood $l_m(\boldsymbol{\beta}_n)$, gradient $\boldsymbol{g}_m(\boldsymbol{\beta}_n)$, and Hessian matrix $\boldsymbol{H}_m(\boldsymbol{\beta}_n)$ using Equations (7) - (9). Then sum the stratum-specific contributions $l_m(\boldsymbol{\beta}_n)$, $\boldsymbol{g}_m(\boldsymbol{\beta}_n)$, and $\boldsymbol{H}_m(\boldsymbol{\beta}_n)$ to obtain the log-likelihood $l(\boldsymbol{\beta}_n)$, gradient $\boldsymbol{g}(\boldsymbol{\beta}_n)$, and Hessian matrix $\boldsymbol{H}(\boldsymbol{\beta}_n)$ using Equation (6).

3) At the analysis center, construct the SSCP matrix as shown in Equation (13) using $\mathbf{A} = \mathbf{I}(\boldsymbol{\beta}_n) = -\boldsymbol{H}(\boldsymbol{\beta}_n)$ and $\mathbf{c} = \boldsymbol{g}(\boldsymbol{\beta}_n)$ and solve the system of linear equations of the form in Equation (12) using PROC REG as described above. This involves a series of steps implemented in the utility macro %***solve_linear_equations_reg*** (part of the package for the analysis center). The macro takes datasets with $\mathbf{I}(\boldsymbol{\beta}_n)$ and $\boldsymbol{g}(\boldsymbol{\beta}_n)$ as inputs, constructs the appropriate SSCP-type dataset, feeds it into PROC REG, and outputs two datasets: one with the solution to the Newton-Raphson equation for $\boldsymbol{b}_n = \boldsymbol{\beta}_{n+1} - \boldsymbol{\beta}_n$ and one containing the inverse matrix $\mathbf{I}^{-1}(\boldsymbol{\beta}_n)$. The latter is only used in the final iteration to estimate the covariance matrix.

4) Repeat steps 1 to 3 until convergence is achieved at the iteration $n+1$ that meets the convergence criterion, $\widehat{\boldsymbol{\beta}} = \boldsymbol{\beta}_{n+1}$

After convergence is achieved, an additional iteration of steps 1 to 4 is necessary to calculate the covariance matrix of parameter estimates $\widehat{cov}(\widehat{\boldsymbol{\beta}}) = \mathbf{I}^{-1}(\boldsymbol{\beta}_{n+1})$. The additional iteration is required because at iteration $n$ we do not know the matrix $\mathbf{I}(\boldsymbol{\beta}_{n+1})$, we only know the matrix $\mathbf{I}(\boldsymbol{\beta}_n)$.

## 2.3 Distributed Cox regression convergence criteria

We use the relative convergence criteria identical to the SAS relative convergence criteria specified by option XCONV. Let $\beta_{n+1,s}$ be the estimate of the parameter $s = 1, \dots, p$ at iteration $n$. The regression criterion is satisfied if:

$$\max_s |\delta_{n+1,s}| < xconv\_value$$

where

$$\delta_{n+1,s} = \begin{cases} \beta_{n+1,s} - \beta_{n,s}, & |\beta_{n,s}| < 0.01 \\ \frac{\beta_{n+1,s} - \beta_{n,s}}{\beta_{n,s}}, & else \end{cases}$$

## 2.4 Goodness-of-fit measures, survival function, and residuals

Our Cox DRA application calculates the following goodness-of-fit measures: log-likelihood, Akaike information criterion (AIC), and Bayesian information criterion (BIC). These measures can be evaluated exactly using summarized data transferred to the analysis center. Additional, quantities of interest are computed on the individual-level data within each data partner site only. These data partner-specific quantities include an estimate of the survival function and martingale and deviance residuals. In **Appendix B**, we give explicit expressions for these measures in the case of horizontally partitioned data.

## 3. How to use the DRA application for distributed Cox regression

### 3.1 Overview

In our companion paper (Her et al. Submitted), we describe the steps involved in setting up and executing the linear and logistic DRA application (see Section 2 "How to use the DRA application"). Many of the steps are the same for Cox DRA. Below we briefly summarize these common steps and then describe what is different for the Cox regression, namely the structure of the input dataset, and the parameters specified in the main macro %*distributed_regression*, and output tables.

The DRA application comprises two interlinked packages of macros and programs – one for the analysis center and one for the data partners. These packages can be downloaded from https://www.sentinelinitiative.org/sentinel/methods/utilizing-data-various-data-partners-distributed-manner. Each package has a master wrapper program: *run_d_reg_central_tmpl.sas* in the package for the analysis center and *run_d_reg_dp_templ.sas* in the package for data partners. The user needs to update a few site-specific parameters in these wrapper programs, as described in Section 3 of the companion paper (Her et al. Submitted). In addition, the analysis center needs to specify a call to the main macro %*distributed_regression* in the wrapper *run_d_reg_central_tmpl.sas*. All regression-related parameters, including the dependent and independent variables, convergence criteria, the list of participating data partners, and type of regression (e.g. Cox) are specified in this macro.

After necessary parameters are updated in the SAS wrappers, users at the analysis center and data partners can execute their SAS programs at a mutually agreed time window. All programs run continuously and exchange data between the analysis center and data partners using the mechanism described in (Her et al. Submitted). The process stops when either the regression algorithm converges, reaches a pre-specified maximum number of iterations, or the program catches an error in the process.

### 3.2 Example dataset

We used the publicly available Maryland State Prison dataset to illustrate the steps involved in using the DRA application for Cox regression (Rossi and Henry 1980). We randomly partitioned the dataset into sizes of $n_1 = 134$, $n_2 = 149$, and $n_3 = 149$. Time to re-incarceration in weeks (*week*) was the time-to-event outcome, and financial aid (*fin*, a binary variable), age (*age*, a continuous variable), and number of prior convictions (*prio*, a continuous variable) were the independent covariates. The censoring variable *arrest* has value of 1 if arrested and 0 otherwise. We added a variable *dp_cd* that served as the data partner identifier that had a value of 1, 2, or 3 in each dataset. This variable was used as the stratification variable in one of the examples in Section 3.3.

### 3.3 Examples of using macro %*distributed_regression* for Cox model

In this section, we provide some examples of how to use %*distributed_regression*, the main macro at the analysis center. The parameters explained below should be sufficient for most practical applications of Cox DRA. The complete list of all parameters and their descriptions can be found in **Appendix C**. Below we use the dataset described in Section 3.2.

*Example 1.* The code below performs Cox DRA using the Breslow approximation without any stratification variables. In addition to required parameters, it specifies three optional parameters *tbl_inital_est, tbl_events_time_set* and *xconv*.

```
/*
  Parameter RunID specifies an identifier for a given macro call. It is used to form a prefix
%let prefix=&RunID._ for all output datasets names.
  Parameter dp_cd_list specifies a list of data partner sites participating in the current request.
  Parameter reg_ds_in specifies the name of the input dataset for regression at a data partner site,
the name and structure the same at all sites. The dataset must be located in the SAS library
data_in defined in the data partner master wrapper program.
  Parameters dependent_vars and independent_vars specify dependent and independent variables
for the regression.
  Parameter censoring_var specifies the censoring variable.
  Parameter regr_type_cd defines the type of the regression: 1- linear; 2- logistic; 10- Cox;
  Parameter tbl_intial_est names the table with initial guesses (starting values) for the regression
parameter estimates. Must have a column for each of the parameter estimates which has the same
name as the corresponding covariate. It should be located in the SAS library named infolder. In
the example below dataset Cox_Model_Coeff_0 has all initial values equal to 0.
  Parameter tbl_events_time_set names the table with all values of event times from all data
partners. The dataset must have one column with the same name as dependent variable. It should
be located in the SAS library named infolder. This is an optional parameter but specifying it can
reduce the number of required iterations.
  Parameter xconv specifies relative convergence criteria.
*/
 %distributed_regression(RunID=dc1
                        ,dp_cd_list=1 2 3
                        ,reg_ds_in=RECID_DR
                        ,dependent_vars=week
                        ,independent_vars=fin age  prio
                        ,censoring_var=arrest
                        ,regr_type_cd=10
                        ,tbl_intial_est=Cox_Model_Coeff_0
                        ,tbl_events_time_set=RECID_Events_Time_Set
                        ,xconv=1e-4
                        ) ;
```

***Example 2.*** This example performs Cox DRA stratified by a data partner identifier, *strata_vars=dp_cd*, using the Efron approximation for ties, instead of the default Breslow approximation. All other parameters are as described in Example 1.

```
/*
Parameter strata_vars specifies the list of stratification variables.
Parameter ties specifies the method of handling ties in event times in Cox regression.
*/

%distributed_regression(RunID=dc2
                        ,dp_cd_list=1 2 3
                        ,reg_ds_in=RECID_DR
                        ,dependent_vars=week
                        ,independent_vars=fin age  prio
                        ,censoring_var=arrest
                        ,regr_type_cd=10
                        ,strata_vars=dp_cd
                        ,ties=EFRON
                        ) ;
```

### 3.4 Creation of output tables

The macro %*distributed_regression* creates final output datasets in the *msoc* subdirectory of the request directory at the analysis center. All datasets from a given execution of the macro have the same prefix determined by the parameter *RunID*. For Cox DRA (*regr_type_cd=10*) the structure of most of these output datasets mirrors corresponding datasets generated by PROC PHREG. The complete list of output datasets and their description are given in **Appendix D**. The output can be generated by printing the output tables in the *msoc* subdirectory or by using the macro %*create_cox_reg_rpt* included with the package at the analysis center. An example of how to use this macro is shown in the wrapper template *run_d_reg_central_tmpl.sas*.

### 4. Example output created by macro %*distributed_regression*

In addition to developing and testing Cox DRA in the Maryland State Prison dataset, we also tested our Cox DRA application on several datasets including publicly available data, simulated data, and empirical datasets from three data partners in the Sentinel System. Examples of the full report generated by the macro %*create_cox_reg_rpt* can be found online at
https://www.sentinelinitiative.org/sentinel/methods/utilizing-data-various-data-partners-distributed-manner.

### 4.1 Main results from distributed Cox regression
In this section, we report the parameter estimates, standard errors, and some goodness-of-fit-measures produced by the main macro %*distributed_regression* for the two examples described in Section 3.3. The results are presented in **Table** 1 to **Table 6**.

| | Total | Event | Censored | Percent Censored |
|---|---|---|---|---|
| | 432 | 114 | 318 | 73.61 |

**Table 1.** Number of events and censored values from Example 1 described in Section 3.3. No stratification by data partner.

| Criterion | Without Covariates | With Covariates |
|---|---|---|
| -2Log L | 1351.366779 | 1322.465221 |
| AIC (smaller is better) | 1351.366779 | 1328.465221 |
| SBC (smaller is better) | 1351.366779 | 1336.673816 |

**Table 2.** Model fit statistics for distributed Cox regression from Example 1 described in Section 3.3. No stratification by data partner. Breslow approximation for ties.

| Parameter | DF | Parameter Estimate | Standard Error | Chi-Square | Pr > Chi-Square | Hazard Ratio | Lower 95% CL Hazard Ratio | Upper 95% CL Hazard Ratio |
|---|---|---|---|---|---|---|---|---|
| fin | 1 | -0.346444 | 0.190236 | 3.316518 | 0.0686 | 0.707198 | 0.4870936 | 1.0267629 |
| age | 1 | -0.066921 | 0.020840 | 10.311876 | 0.0013 | 0.935269 | 0.8978378 | 0.9742614 |
| prio | 1 | 0.096528 | 0.027241 | 12.556144 | 0.0004 | 1.101341 | 1.0440804 | 1.1617414 |

**Table 3.** Parameter estimates from Example 1 described in Section 3.3. No stratification by data partner. Breslow approximation for ties.

| Stratum | dp_cd | Total | Event | Censored | Percent Censored |
|---|---|---|---|---|---|
| 1 | 1 | 134 | 36 | 98 | 73.13 |
| 2 | 2 | 149 | 42 | 107 | 71.81 |
| 3 | 3 | 149 | 36 | 113 | 75.84 |
| Total | | 432 | 114 | 318 | 73.61 |

**Table 4.** Number of events and censored values from Example 2 described in Section 3.3. Stratified by data partner.

| Criterion | Without Covariates | With Covariates |
|---|---|---|
| -2Log L | 1100.863717 | 1072.203117 |
| AIC (smaller is better) | 1100.863717 | 1078.203117 |
| SBC (smaller is better) | 1100.863717 | 1086.411712 |

**Table 5.** Model fit statistics for distributed Cox regression from Example 2 described in Section 3.3. Stratification by data partner. Efron approximation for ties.

| Parameter | DF | Parameter Estimate | Standard Error | Chi-Square | Pr > Chi-Square | Hazard Ratio | Lower 95% CL Hazard Ratio | Upper 95% CL Hazard Ratio |
|---|---|---|---|---|---|---|---|---|
| fin | 1 | -0.357741 | 0.191086 | 3.504934 | 0.0612 | 0.699254 | 0.4808197 | 1.0169222 |
| age | 1 | -0.066485 | 0.020877 | 10.141969 | 0.0014 | 0.935677 | 0.8981634 | 0.9747566 |
| prio | 1 | 0.096662 | 0.027419 | 12.428122 | 0.0004 | 1.101488 | 1.043856 | 1.1623017 |

**Table 6.** Parameter estimates for distributed Cox regression from Example 2 described in Section 3.3. Stratification by data partner. Efron approximation for ties.

## 4.2 Residual diagnostics for distributed Cox regression

There are several types of residuals that can be defined for Cox regression. Our DRA application calculates martingale and deviance residuals (see **Appendix** B). To comply with the privacy requirements of data partners, our approach leaves the final individual-level dataset with all initial variables, survival function estimate, and residuals at the data partners, and brings back only summarized results to the analysis center. Each data partner has an option to define a minimum number of records to be summarized (minimum number per cell) by specifying this number in the macro parameter *min_count_per_grp* in their master wrapper program (see template *run_d_reg_dp_templ.sas*). If *min_count_per_grp* is not specified by a data partner, then the parameter *min_count_per_grp_glob*, specified in the macro %***distributed_regression*** is used. Below we illustrate options for residuals using the Maryland State Prison dataset described in Section 3.2 and setting *min_count_per_grp=6*.

To summarize the final output from Cox DRA, we first sort the individual-level output dataset at each data partner site by the linear predictor $\theta = \boldsymbol{\beta}^T \boldsymbol{Z}$. We then group the data into bins based on percentiles of $\boldsymbol{\beta}^T \boldsymbol{Z}$ and calculate the means of the linear predictor and residuals for each bin. Due to ties, the number of observations can vary slightly between bins (Her et al. Submitted). The number of bins in the summary dataset created at the data partner sites can be modified by changing the parameter *groups* in the main macro **%*distributed_regression***. However, if the value of *groups* is too large to satisfy the constraint set by *min_count_per_grp*, the actual number of bins is decreased accordingly by the program. The summarized datasets from each data partner are brought to the analysis center and combined into a single dataset, which can then be used to visually evaluate the goodness-of-fit (see description of the dataset *prefix0.resid_sum_by_pct* in **Appendix D**.

In Figure 1, we provide a plot of the mean martingale residual versus mean linear predictor calculated for each bin for the stratified Cox DRA described in Example 2 (Section 3.3). We used 10 bins (deciles) at each data partner, which is the default value for the parameter *groups*. Random scatter of data points around zero suggests reasonable model fit.

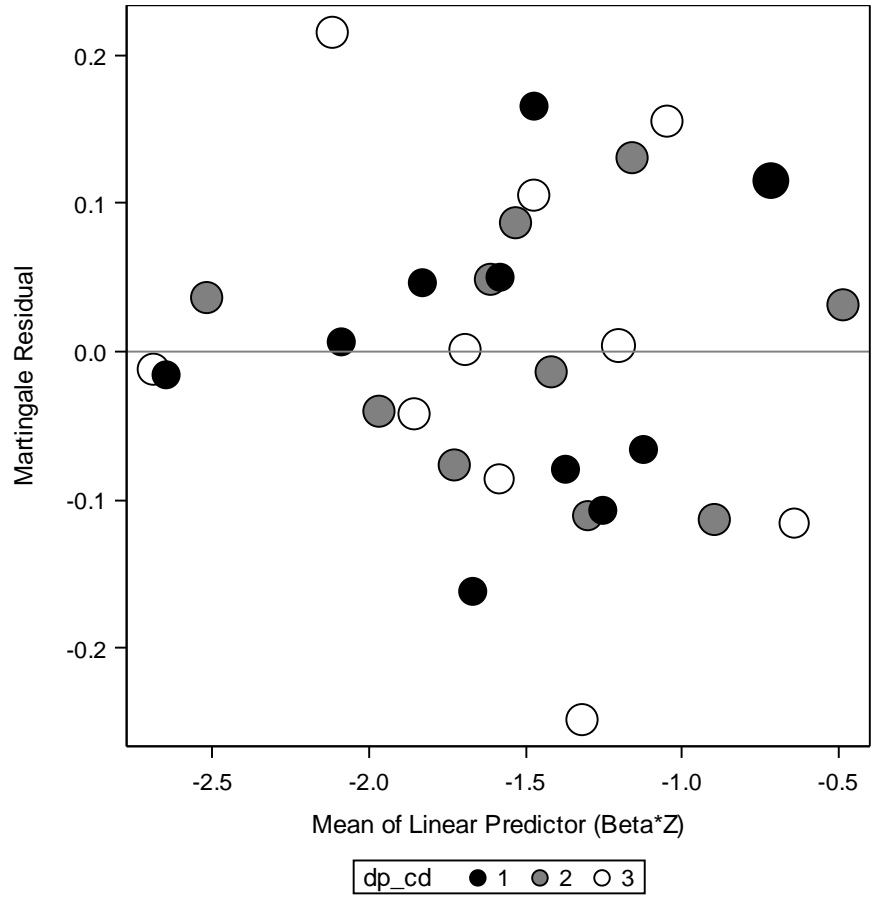

**Figure 1.** Mean martingale residual vs. mean linear predictor for stratified distributed Cox regression from Example 2 described in Section 3.3. The data are grouped into bins by decile of the linear predictor $\theta = \beta^T Z$. The radius of each bubble is proportional to the number of observations in a bin. Different symbols represent data points from different data partners. If the model is correctly specified then the data points are expected to scatter randomly around zero.

### 4.3 Comparison of the results for distributed Cox regression with results obtained using standard SAS procedures on pooled individual-level dataset

We tested our Cox DRA application on several datasets (Sentinel System 2018). In all our tests, DRA algorithms produced regression parameters and standard error estimates that were in complete agreement with the results produced by standard SAS procedures on the combined data. Specifically, they are within machine precision (1E-16) when we used the same input parameters including initial guesses for parameters estimates.

In Table 7 and Table 8, we compare results obtained for Examples 1 and 2, respectively, as described in Section 3.3. The results for Cox regression on the combined individual-level dataset were obtained using PROC PHREG.

|  | DR | | Pooled Individual-Level | | | |
| --- | --- | --- | --- | --- | --- | --- |
| Parameter | Parameter Estimate | Standard Error | Parameter Estimate | Standard Error | Difference in Estimates | Difference in Standard Errors |
| fin | -0.346444 | 0.19024 | -0.34644 | 0.19024 | 2.22E-16 | -2.78E-17 |
| age | -0.066921 | 0.02084 | -0.06692 | 0.02084 | -1.39E-16 | 2.78E-17 |
| prio | 0.0965283 | 0.02724 | 0.09653 | 0.02724 | -1.80E-16 | 1.73E-17 |

Table 7. Distributed Cox regression vs. pooled individual-level Cox regression from Example 1 described in Section 3.3. No stratification by data partner. Breslow approximation for ties.

|  | DR | | Pooled Individual-Level | | | |
| --- | --- | --- | --- | --- | --- | --- |
| Parameter | Parameter Estimate | Standard Error | Parameter Estimate | Standard Error | Difference in Estimates | Difference in Standard Errors |
| fin | -0.357741 | 0.19109 | -0.35774 | 0.19109 | 2.22E-16 | 0.00E+00 |
| age | -0.066485 | 0.02088 | -0.06649 | 0.02088 | 5.55E-17 | -7.63E-17 |
| prio | 0.0966619 | 0.02742 | 0.09666 | 0.02742 | -5.55E-17 | 1.04E-17 |

Table 8. Distributed Cox regression vs. pooled individual-level Cox regression from Example 2 described in Section 3.3. Stratification by data partner. Efron approximation for ties.

## 5. Discussion

We have successfully developed, tested, and validated a DRA application for Cox proportional hazards regression. The application requires only Base SAS and SAS/STAT modules and can be used on any operating system on which SAS can be installed (Windows, Unix, or Linux, etc.). Our Cox DRA application produces results identical, with machine precision, to the results obtained from the corresponding pooled individual-level data analysis with the PROC PHREG procedure. The application can implement either the Breslow or Efron approximation for tied events and can fit either a non-stratified or stratified Cox model. In the context of DRA within DDNs, a stratified Cox model by data partner is strongly preferred. It allows a site-specific baseline hazard function, which is often more realistic in multi-database studies. It also allows for greater summarization of data shared between data partners and the analysis center, reducing both risks to privacy and overall execution time. Our DRA application uses a modified computational algorithm when the model is stratified by data partner and does as much as possible to summarize the data shared by data partners with the analysis center.

We integrated our DRA application for Cox regression into PopMedNet, an open-source distributed networking software that allows automatable iterative file transfer between data partners and the analysis center (Her et al. 2018). An important advantage of using PopMedNet is that all file transfers between data partners and the analysis center are achieved through standard secure HTTPS/SSL/TLS connections. There is no external access to data partner data. Both PopMedNet and SAS processes are run under the user accounts that the data partners create and maintain. While our current implementation of DRA uses PopMedNet, our DRA application can be implemented manually or with any data transferring software that meets the specifications described in our companion paper (Her et al. Submitted).

We plan to extend our Cox DRA algorithm to include calculation of the robust sandwich covariance estimator and robust standard errors. Future work will expand the Cox DRA application to vertically partitioned data environments, where different databases contain information from the same individuals (Li et al. 2016, Fienberg et al. 2006, Reiter et al. 2004).

# Appendix

## A. Efron approximation for Cox distributed regression analysis in the case of horizontally partitioned data.

The Efron approximation provides a correction to the Breslow approximation when there are a relatively large number of ties. For this approximation, in addition to site-specific matrices $d_{m,j,k}^{(l)}(\boldsymbol{\beta}_n)$ and $S_{m,j,k}^{(l)}(\boldsymbol{\beta}_n)$ defined in the context of the Breslow approximation in Section 2.2, one also needs to also calculate the following matrices:

$$Q_{m,j,k}^{(l)}(\boldsymbol{\beta}) = \sum_{i=1}^{n_{m,k}} I(T_{i,m,k} = t_{m,j}, \Delta_{i,m,k} = 1) w_{i,m,k} \exp(\boldsymbol{\beta}^T Z_{i,m,k}) Z_{i,m,k}^l \quad (17)$$

$$Q_{m,j,k}^{(l)}(\boldsymbol{\beta}) = \sum_{i=1}^{n_{m,k}} I(T_{i,m,k} = t_{m,j}, \Delta_{i,m,k} = 1) w_{i,m,k} \exp(\boldsymbol{\beta}^T Z_{i,m,k}^l)$$

where, for $l = 0,1,2$, $Z_{i,m,k}^0 = 1$, $Z_{i,m,k}^1 = Z_{i,m,k}$, $Z_{i,m,k}^2 = Z_{i,m,k} Z^T_{i,m,k}$. such that $Q_{m,j,k}^{(l)}(\boldsymbol{\beta})$ is a scalar when $l = 0$, a vector with dimension $p$ when $l = 1$ and a matrix with dimensions $p * p$ when $l = 2$.

In the formulas below, a variable without a subscript implies a sum over that subscript. In particular, the absence of the $k$ index implies summation over all data partners. For example:

$$Q_{m,j}^{(l)}(\boldsymbol{\beta}) = \sum_k Q_{m,j,k}^{(l)}(\boldsymbol{\beta}) \quad (18)$$

We also define the following additional matrices:

$$S_{m,j,k,s}^{(l,E)}(\boldsymbol{\beta}) = S_{m,j,k}^{(l)}(\boldsymbol{\beta}) - \frac{s-1}{d_{m,j}} Q_{m,j,k}^{(l,E)}(\boldsymbol{\beta}) \quad (19)$$

Using these matrices and $d_{m,j,k}^{(l)}$ from Equation (3) the partial log-likelihood $l^{(E)}(\boldsymbol{\beta})$, gradient vector $g^{(E)}(\boldsymbol{\beta})$, and the Hessian matrix $H^{(E)}(\boldsymbol{\beta})$ can be calculated under the Efron approximation as follows:

$$l^{(E)}(\boldsymbol{\beta}) = \sum_{m=1}^{M} l_m^{(E)}(\boldsymbol{\beta})$$
$$g^{(E)}(\boldsymbol{\beta}) = \sum_{m=1}^{M} g_m^{(E)}(\boldsymbol{\beta})$$
$$H^{(E)}(\boldsymbol{\beta}) = \sum_{m=1}^{M} H_m^{(E)}(\boldsymbol{\beta}) \quad (20)$$

$$l_m^{(E)}(\boldsymbol{\beta}) = \sum_j \left\{ \boldsymbol{\beta}^T \boldsymbol{d}_{m,j}^{(1)} - \frac{d_{m,j}^{(0)}}{d_{m,j}} \sum_{s=1}^{d_{m,j}} \log S_{m,j,s}^{(0,E)}(\boldsymbol{\beta}) \right\} \quad (21)$$

$$\boldsymbol{g}_m^{(E)}(\boldsymbol{\beta}) = \sum_j \left\{ \boldsymbol{d}_{m,j}^{(1)} - \frac{d_{m,j}^{(0)}}{d_{m,j}} \sum_{s=1}^{d_{m,j}} \frac{\boldsymbol{S}_{m,j,s}^{(1,E)}(\boldsymbol{\beta})}{S_{m,j,s}^{E}(\boldsymbol{\beta})} \right\} \quad (22)$$

$$\boldsymbol{H}_m^{(E)}(\boldsymbol{\beta}) = -\sum_j \frac{d_{m,j}^{(0)}}{d_{m,j}} \sum_{s=1}^{d_{m,j}} \left\{ \frac{\boldsymbol{S}_{m,j,s}^{(2,E)}(\boldsymbol{\beta})}{S_{m,j,s}^{E}(\boldsymbol{\beta})} - \frac{\boldsymbol{S}_{m,j,s}^{(1,E)}(\boldsymbol{\beta}) * \left[\boldsymbol{S}_{m,j,s}^{(1,E)}(\boldsymbol{\beta})\right]^T}{\left[S_{m,j,s}^{E}(\boldsymbol{\beta})\right]^2} \right\} \quad (23)$$

Note that the only difference between the scalar $d_{m,j}^{(0)}$ and the number of events $d_{m,j}$ used in the above equations is that the former is calculated using weights (see Equation 3). When $w_{i,m,k} = 1$ these quantities are the same for all subjects.

The main steps of our DRA computational algorithm using the Efron approximation are similar to the ones described in the context of the Breslow approximation in Section 2.2. The only difference is that one needs to use matrices $\boldsymbol{S}_{m,j,k,s}^{(l,E)}(\boldsymbol{\beta})$ instead of matrices $\boldsymbol{S}_{m,j,k}^{(l)}(\boldsymbol{\beta})$ and perform an additional summation over the index $s$ when calculating the log-likelihood, gradient, and Hessian matrix using equations (21) – (23).

B. **Expressions for goodness-of-fit measures, survival function, and residuals in the case of distributed data.**

Akaike Information Criterion is defined as:

$$AIC = -2l(\hat{\boldsymbol{\beta}}) + 2p \qquad (24)$$

Bayesian Information Criterion is defined as:

$$BIC = -2l(\hat{\boldsymbol{\beta}}) + p \ln(\sum_{i,m,k} \Delta_{i,m,k}) \qquad (25)$$

Estimators for cumulative baseline hazard function (minus log of baseline survival function).

Breslow estimator:

$$h_{cum}^{(m,0)}(T) = \sum_j \frac{d_{m,j}}{S_{m,j}} I(T \geq t_{m,j}) \qquad (26)$$

Fleming-Harrington Estimator for Efron approximation:

$$h_{cum}^{(m,0)}(T) = \sum_j \sum_{s=1}^{d_{j,m}} \frac{1}{S_{m,j,s}^{(E)}} I(T \geq t_{m,j}) \qquad (27)$$

Note that both cumulative baseline hazard estimators change only at event times $T = t_{m,j}$ and stay constant in between event times.

Cumulative hazard function (minus log of survival function):

$$h_{cum}^m(T_{i,m,k}, \boldsymbol{\beta}^T \mathbf{Z}_{i,m,k}) = \exp(\boldsymbol{\beta}^T \mathbf{Z}_{i,m,k}) \, h_{cum}^{(m,0)}(T_{i,m,k}) \qquad (28)$$

Survival function:

$$S_{surv}^m(T_{i,m,k}) = \exp\left(-h_{cum}^m(T_{i,m,k}, \boldsymbol{\beta}^T \mathbf{Z}_{i,m,k})\right) \qquad (29)$$

Martingale residuals:

$$M_{i,m,k} = \Delta_{i,m,k} - h^m_{cum}(T_{i,m,k}, \boldsymbol{\beta}^T \boldsymbol{Z}_{i,m,k}) \tag{30}$$

Deviance residuals:

$$D_{i,m,k} = sign(M_{i,m,k})\sqrt{2[-M_{i,m,k} - \Delta_{i,m,k} log(\Delta_{i,m,k} - M_{i,m,k})]} \tag{31}$$

It is useful to plot the martingale or deviance residuals against the linear predictor $\boldsymbol{\beta}^T \boldsymbol{Z}_{i,m,k}$. The random scatter plot around 0 with no trend indicates that the chosen functional form for the log of hazard as a function of independent variables is reasonable.

## C. Parameters for the main macro *%distributed_regression*.

The table below describes all parameters for the macro *%distributed_regression*.

| Parameter | Description |
| --- | --- |
| RunID | Identifier for a given macro call. It is used to form a prefix &RunID for the names of all output datasets. This allows calling this macro several times within the same distributed regression request. Preferably less than 4 characters. *Required.* *Example: RunID=dr1* |
| reg_ds_in | The name of the input analytic dataset. The dataset must have the same name at all data partner sites and located in the SAS library called DATA_IN (defined in the data partner's SAS wrapper). *Required.* *Example: reg_ds_in=LINEAR_KARR_2005* |
| dp_cd_list | The list of participating data partners separated by space. *Required.* *Example: dp_cd_list=7 15 19* |
| regr_type_cd | Defines the type of the regression. 1=linear; 2=logistic; 10=Cox. *Required.* *Example: regr_type_cd=1* |
| dependent_vars | Name of the dependent variable in the regression. *Required.* *Example: dependent_vars= medv* |
| independent_vars | List of the independent variables in the regression. *Required.* *Example: independent_vars=crim indus dis* |
| NOINT | When set to NOINT the regression analysis fits without an intercept. Default is blank, which fits the model with an intercept. Not relevant for Cox regression. *Optional.* |
| Freq | Name of the variable that indicates frequency of an observation. *Optional.* *Example: freq=freq_variable* |

| Weight | Name of the variable that indicates weight of an observation. *Optional.* *Example, weight=weight_variable* |
|---|---|
| tbl_intial_est | Name of the table with initial guesses of the regression parameter estimates. This table must have a column for each of the parameter estimates. The column names for the parameter estimates should be the same as the names of corresponding independent variables specified in *independent_vars*. (Same structure as special SAS dataset of the TYPE=PARM). If *tbl_inital_est* is not specified, all initial guesses are set to 0. Note, this is different from the default in PROC LOGISTIC, which sets all initial guesses to 0 except for the intercept, which is set to the average of the outcome variable. The dataset *tbl_intial_est* should be located in the SAS library infolder defined in the wrapper for the analysis center. Not relevant for linear regression. *Optional.* *Example: tbl_intial_est=Model_Coeff_0* |
| xconv | Relative convergence criteria. The same definition as the one used by SAS. *Optional.* Default is 1E-4. |
| max_iter_nb | Maximum number of allowed iterations. *Optional*. Default is 20. |
| censoring_var | Name of the censoring variable in the Cox regression. Required for Cox regression. Not relevant for other types of regression. *Example: censoring_var=arrest* |
| censoring_lev | Value of the censoring variable indicating censored observations in the Cox regression. Not relevant for other types of regression. *Optional*. Default is 0. |
| strata_vars | Name of the list of stratification variables for Cox regression. Default is blank (no stratification). When the list of stratification variables includes the data partner identifier variable *dp_cd*, the calculation is done much more efficiently. In this case, all summations over event times are pushed to the data partner resulting in much smaller dataset transfer to the analytic center. *Optional* *Example: strata_vars=dp_cd, sex.* |
| ties | Specifies the method of handling ties in event times in Cox regression. Not relevant for other types of regression. Allowed values are: BRESLOW or ERFRON. *Optional*. Default: *Example: ties=Breslow*. |
| tbl_events_time_set | Name of the table with all possible value of event times at each data partner (no censoring times). Must have one column with the same name as the dependent variable. If not specified then *tbl_events_time_set* is created at the analysis center after the first iteration. The parameter is ignored when stratification variables include variable *dp_cd* (an identifier for a data partner). *Optional*. *Example: tbl_events_time_set= events_time_set* |
| alpha | Level of statistical significance. *Optional*. Default is 0.05. |
| groups | Number of groups used in the calculation of residuals summary statistics. *Optional*. Default is 10. |
| wait_time_min | Minimum time interval for checking for the trigger file *files_done.ok*. Measured in seconds. Used by the macro *%file_watcher*. *Optional*. Default is 3. |
| wait_time_max | Maximum time interval for checking for the trigger file *files_done.ok*. Measured in seconds. Used by the macro *%file_watcher*. *Optional*. Default is 7,200, which is 2 hours. |
| last_runid_in | If one wants to run more than one regression (can be different regression models) within the same request one should specify *last_runid_in=0* for the first few calls of this macro and to 1 for the last call. *Optional*. Default is 1. |
| test_env_cd | Set to 1 to execute the DRA application in the special development/testing environment. The directory structure in this environment is the same as the structure at the analysis center. When set to 1, the program can be executed within a single SAS session with the code for different data partners running sequentially. It allows testing of most of the SAS code without the need of a data transferring software. *Optional*. Default is 0 which means production environment. |

| max_numb_of_grp | Sets upper limit to the number groups for summarized data returned to the analysis center. Normally the number of groups is determined by parameters *min_count_per_grp* or *min_count_per_grp_glob*. However, for large datasets this can result in large amount of data transferred from data partners to the analysis center. This is often unnecessary and this parameter puts a cap on the number of rows returned to the analysis center. *Optional*. Default is 10,000. |
|---|---|
| min_count_per_grp_glob | Sets minimum count per cell for summarized data returned to the analysis center. It is only used if a data partner site does not specify parameter *min_count_per_grp* in their master program. This affects datasets used for residual analysis and goodness-of-fit measures (ROC and Hosmer-Lemeshow statistic for logistic regression). *Optional*. Default is 6. |

## D. Final output datasets

Below is a table with the list and description of final output datasets created by the macro %*distributed_regression* for Cox regression. All datasets are located in the subdirectory *msoc* at the analysis center. The datasets from a given run have the same prefix equal *&RunID*. For example, for *&RunID=dc1* the *&prefix=dc1*.

| Dataset Name | Dataset Description |
|---|---|
| &PREFIX.BASELN_HAZARD | The dataset has an estimate of cumulative baseline hazard function at the event times of each stratum. The non-stratified case is treated as the case with single strata. |
| &PREFIX.BASELN_SURVIVAL | The dataset has an estimate of the survival function at the event times of each stratum. The survival function is evaluated at average values of the covariates per stratum. The non-stratified case is treated as the case with single strata. |
| &PREFIX.CENS_SUM | Has similar structure as the ODS table *CensoredSummary* generated by PROC PHREG. It has summary of event and censored observations. |
| &PREFIX.CONVRG_STATUS | Has similar structure as the ODS table *ConvergenceStatus* generated by PROC PHREG. Also contains information about number of iteration and convergence criteria. |
| &PREFIX.COV_EST | Has the same structure as the ODS table *CovB* generated by PROC PHREG. Includes information about model-based covariance of estimates. |
| &PREFIX.GLOB_NULL_CHISQ | Has the same structure as the ODS table *GlobalTests* generated by PROC PHREG. Includes Likelihood Ratio, Chi-Square statistic, degrees of freedom, and p-value for the global null hypothesis test. |
| &PREFIX.ITER_PARMS_HIST | Has the same structure as the ODS table *IterHistory* generated by PROC PHREG. It includes information about iteration history. |

| Dataset Name | Dataset Description |
|---|---|
| &PREFIX.MODELFIT | Has structure similar to the ODS table *ModelFit* generated by PROC PHREG. It has information about various goodness-of-fit measures including -2Log L (L is likelihood), AIC, and BIC criteria. |
| &PREFIX.MODELINFO | Has the same structure as the ODS table *ModelInfo* generated by PROC PHREG. It includes information about names of the input dataset, dependent variable, censoring variable, and type of approximation for handling event ties (Breslow or Efron). |
| &PREFIX.MODEL_COEFF | Has the similar structure as the output dataset specified by option *OUTEST* in PROC PHREG. Includes information about regression coefficient in the longitudinal form: a single row with a column for each of the coefficient. |
| &PREFIX.P_EST | Has the same structure as the ODS table *ParameterEstimates* generated by PROC PHREG. Includes information about regression coefficient in the vertical form with separate row for each coefficient. In addition, it has columns for model standard errors, p-values, upper and lower confidence limits. |
| &PREFIX.RESID_SUM | The dataset has overall summaries for a number of quantities including number of events, log likelihood, and various goodness-of-fit measures. |

| Dataset Name | Dataset Description |
|---|---|
| &PREFIX.RESID_SUM_BY_PCT | Has summary statistics based on final output dataset at each data partner. The data are grouped by percentiles of the linear predictor values $\theta_j = \boldsymbol{\beta}^T \boldsymbol{Z}_j$. The number of observation per bin for a data partner can vary slightly due to ties. The number of bin is determined by the macro parameter *groups* specified in the main macro %*distributed_regression*. The default value is *groups=10*. The summary statistics include mean values of linear predictor, martingale and deviance residuals. It also includes data partner identifier variable *dp_cd* and a number of observations per bin. The dataset can be used to generate plots of martingale and deviance residuals and visually evaluate the goodness-of-fit by the regression model. |